\documentclass{nature}

\usepackage{graphicx}

\usepackage{epstopdf}
\usepackage{amsbsy}
\usepackage{appendix}
\usepackage{ulem}
\usepackage{color}
\usepackage{gensymb}
\usepackage{amsmath}
\usepackage{amsthm}
\usepackage{soul}

\newcommand{\beq}{\begin{equation}}
\newcommand{\eeq}{\end{equation}}
\newcommand{\beann}{\begin{eqnarray*}}
\newcommand{\eeann}{\end{eqnarray*}}

\definecolor{ao(english)}{rgb}{0.0, 0.5, 0.0}

\title{Exact Self-Consistent Effective Hamiltonian Theory}

\author{Xindong Wang$^{1}$, Xiao Chen$^{2}$, Liqin Ke$^{3}$, Hai-Ping Cheng$^{2}$, B. N. Harmon$^{3}$}

\begin{document}
\clearpage\maketitle
\thispagestyle{empty}

\begin{affiliations}
\item  Sophyics Technology, LLC
\item  Department of Physics and the Quantum Theory Project, University of Florida, Gainesville, Florida 32611, USA
\item  Ames Laboratory, U.S.~Department of Energy, Ames, Iowa 50011
\end{affiliations}

\date{\today}

\begin{abstract}
We propose a general variational fermionic many-body wavefunction that generates an effective Hamiltonian in a quadratic form, which can then be exactly solved. The theory can be constructed within the density functional theory framework, and a self-consistent scheme is proposed for solving the exact density functional theory. We apply the theory to structurally-disordered systems, symmetric and asymmetric Hubbard dimers, and the corresponding lattice models. The single fermion excitation spectra show a persistent gap due to the fermionic-entanglement-induced pairing condensate. For disordered systems, the density of states at the edge of the gap diverges in the thermodynamic limit, suggesting a topologically ordered phase. A sharp resonance is predicted as the gap is not dependent on the temperature of the system. For the symmetric Hubbard model, the gap for both half-filling and doped case suggests that the quantum phase transition between the antiferromagnetic and superconducting phases is continuous.

\end{abstract}

\clearpage
\setcounter{page}{1}

Exact solutions to interacting many-body fermionic systems, especially in dimensions higher than one, have always been the holy grail for quantum physicists. Computationally, the exponential growth of the Fock-Hilbert space's dimension with the size of the system (number of fermions, for example) also makes it a benchmark problem for quantum computations. In this paper, we propose an exact variational wavefunction approach based on the non-empty nature of the renormalized fermion vacuum state. 

The present paper is organized in the following way. In Section 1, the renormalized fermion vacuum state is introduced as a Bogolyubov transformation of the original fermion creation and annihilation operators, and by constraining the number of local basis for the renormalized ground state to three for each onsite band, we proposed a dynamical mechanism for the spontaneous time reversal symmetry. In Section 2, we present the many-body variational wavefunction in the framework of Density Functional Theory~\cite{hohenberg1964pr} and the single fermion excitations are solutions of the quadratic effective Hamiltonian. In Section 3, we apply the theory to quantum proton transport in liquid water and show that the fermionic nature of the proton exchange in Hydrogen bonds in water~\cite{Lowdin1966pr} will drive the system into a topologically ordered state. A sharp electric resonance at low frequency is predicted. In Section 4, we apply the theory to Hubbard dimer and A-B lattice problem and show that a persistent gap opens up at around the chemical potential of the single fermions. The application of the theory is clearly not confined to these two examples, and can be applied to all fermionic many-body systems.

\section{Vacuum Engineering and Time Reversal Symmetry Breaking}
For any fermionic system, we can always decompose the full Hilbert space into the tensor product of the two sub-Hilbert space, $S\bigotimes E$, where $S$ is a single band fermionic system at spatial location ${\bf x}$, and $E$ is the rest of the system. For each ${\bf x}$, we have the following  complete set of orthonormal local states, defined by the fermion creation operators
\begin{equation}
    |0\rangle,  \  \ \hat{\psi}^{\dagger}_{\uparrow} ({\bf x}) |0\rangle, \ \ \hat{\psi}^{\dagger}_{\downarrow} ({\bf x}) |0\rangle, \  \ \hat{\psi}^{\dagger}_{\uparrow} ({\bf x})\hat{\psi}^{\dagger}_{\downarrow} ({\bf x}) |0\rangle,
\end{equation}

We define the following Bogolyubov vacuum at ${\bf x}$ as
\begin{equation}
    |Vac({\bf x})\rangle = \alpha({\bf x}) |0\rangle + \beta({\bf x})  \hat{\psi}^{\dagger}_{\uparrow} ({\bf x})\hat{\psi}^{\dagger}_{\downarrow} ({\bf x}) |0\rangle \\
\end{equation}
where $\alpha({\bf x})$ and $\beta({\bf x})$ are c-numbers and 
\begin{equation}
|\alpha({\bf x})|^2 + |\beta({\bf x})|^2 = 1, \   \  2 |\beta({\bf x})|^2 = n_{Vac}({\bf x})
\end{equation}
where $n_{Vac}({\bf x})$ is the density of particle at ${\bf x}$ of the non-empty Bogolyubov vacuum.
And we further define the following Bogolyubov transformed operators
\begin{equation}
\begin{aligned}
    \hat{\xi}_{\downarrow}({\bf x}) &=& \beta({\bf x}) \hat{\psi}^{\dagger}_{\uparrow}({\bf x})+\alpha({\bf x}) \hat{\psi}_{\downarrow}({\bf x}) \\
    \hat{\xi}^{\dagger}_{\downarrow}({\bf x}) &=& \beta^*({\bf x}) \hat{\psi}_{\uparrow}({\bf x})+\alpha^*({\bf x}) \hat{\psi}^{\dagger}_{\downarrow}({\bf x})\\
    \hat{\xi}_{\uparrow}({\bf x}) &=& \beta({\bf x}) \hat{\psi}^{\dagger}_{\downarrow}({\bf x})-\alpha({\bf x}) \hat{\psi}_{\uparrow}({\bf x}) \\
    \hat{\xi}^{\dagger}_{\uparrow}({\bf x}) &=& \beta^*({\bf x}) \hat{\psi}_{\downarrow}({\bf x})-\alpha^*({\bf x}) \hat{\psi}^{\dagger}_{\uparrow}({\bf x})
\end{aligned}
\end{equation}
It is straightforward to verify the following
\begin{equation}
\begin{aligned}
    \hat{\xi}_{\sigma}({\bf x}) |Vac({\bf x})\rangle &= 0\\
    \hat{\xi}^{\dagger}_{\sigma}({\bf x}) |Vac({\bf x})\rangle &= -\sigma \hat{\psi}^{\dagger}_{\sigma}({\bf x})|0\rangle = -\sigma|{\bf x} \sigma\rangle \\
\end{aligned}
\end{equation}
and the following anti-commutation relations
\begin{equation}
    \{\hat{\xi}^{\dagger}_{\sigma}({\bf x}), \hat{\xi}_{\sigma'}({\bf x'})\} = \delta_{{\bf x, x'}}\delta_{\sigma, \sigma'}
\end{equation}
\begin{equation}
\{\hat{\xi}_{\sigma}({\bf x}), \hat{\xi}_{\sigma'}({\bf x'})\} = \{\hat{\xi}^{\dagger}_{\sigma}({\bf x}), \hat{\xi}^{\dagger}_{\sigma'}({\bf x'})\}=0
\end{equation}

We note that the $\xi_\sigma^\dagger ({\bf x})$ creates a fermion from $|Vac({\bf x})\rangle$, with a fractional charge of $|e|(|\alpha({\bf x})|^2 - |\beta({\bf x})|^2)=|e|(1 - 2|\beta({\bf x})|^2)$ and spin $\sigma$. 

The transformation effectively defines a general class of fermions, b-fermions, that turn into electrons when $\beta=0$, positrons when $\alpha=0$, Majorana fermions when $|\beta|=1/2$.

For non-interacting fermionic system, the Bogolyubov transformation is not expected to change the eigenvalues of the system. These transformations correspond to local $U(2)\bigoplus AU(2)$ transformation on the $|Vac({\bf x})\rangle$ state. 

On the other hand, even a small interaction will break the above $U(2)\bigoplus AU(2)$ symmetry between particle vacuum ($|0\rangle$) and hole vacuum ($\hat{\psi}^{\dagger}_{\uparrow} ({\bf x})\hat{\psi}^{\dagger}_{\downarrow} ({\bf x}) |0\rangle$) states. Thus the renormalized (self-consistent) ground state is expected to have fixed $|\alpha({\bf x})|$ and $|\beta({\bf x})|$. 

We note the striking mathematical similarity between the local quantum charge field $[\alpha({\bf x}), \beta({\bf x})]$ and the fermion spin. We will from now call it the espin, as we will show that it gives rise to the quantum electric-dipole moment, or quantum electric moment, an electrical counter part to the magnetic moment which is related to the electron spin. So when the charge at ${\bf x}$ is not $1$, we will have a local effective espin field that fixes the angle between the espin and the effective espin field. The value of the polar angle $\theta({\bf x})$is determined by the local charge as 
\begin{equation}
    \tan(\theta({\bf x})/2) = \frac{|\beta({\bf x})|}{|\alpha({\bf x})|},  \   \  \theta({\bf x})\in [0,\pi]
\end{equation}

These espins can thus be represented by the Pauli matrices with the polar direction, i.e., the eigenstate of the $\sigma_3$ matrix identified to be $|0\rangle$ and  $|\uparrow\downarrow\rangle$ respectively:
\begin{equation}
    \hat{{\bf \sigma}} \cdot {\bf e} \begin{bmatrix}
    \alpha \\ \beta 
    \end{bmatrix} = \begin{bmatrix}
    \alpha \\ \beta 
    \end{bmatrix}
\end{equation}
where ${\bf e}$ is the unit vector in the espin space.

The fermionic hopping terms together with the on-site intra-band effective electric field are expected to be responsible for the ordering of these espins, just the same as the regular spins. 

Without loss of generality, we can also have the following gauge convention
\begin{equation}
    \alpha({\bf x}) =|\alpha({\bf x})|  \     \    \beta({\bf x}) = |\beta({\bf x})|e^{i\varphi({\bf x})}
\end{equation}
From here point on, we will always assume the above gauge convention and assume $\alpha, \beta$ to be positive real numbers and keep the phase field $\varphi({\bf x})$ explicitly. The self-consistent effective Hamiltonian theory is then casting the ground state solution to variational problem in finding $\alpha({\bf x}), \beta({\bf x}), \varphi({\bf x})$ that minimizes the energy. 

In espin space, the net charge is the $\langle\sigma_3\rangle$, and $\hat{\xi}^\dagger, \hat{\xi}$ are the espin raising and lowering operators. The chemical potential can be expressed as 
\begin{equation} \label{particle espin relation}
   -\mu\hat{N} = 2 \mu  \sum_{\bf x, n, \sigma} \sigma_3^{n,\sigma}({\bf x}) 
\end{equation}
i.e., the global espin symmetry is broken with a non-zero chemical potential, i.e., gap in single fermion excitation spectrum. Here, we have explicitly used $n$ to denote that at ${\bf x}$, they may be different bands, in addition to different spins. 

Using the local Bogolyubov vacuum states, we can define the following global Bogolyubov vacuum state
\begin{equation}
    |Vac\rangle = \prod_{\bf x} |Vac({\bf x})\rangle
\end{equation}

The ground state $|Gnd\rangle$ can be viewed as the resulting adiabatic switching on of the coupling term that gives rise to the inter-site couplings between the espins, i.e., the intersite kinetic hopping terms that exchange fermions. The intersite coupling terms will induce the ordering of the espins, thus, the spontaneous time-reversal symmetry breaking in the thermodynamic limit. 

Another way to understand the espin ordering comes from the realization that when the ground state is a degenerate Fermi liquid, each of the degenerate ground states is a Fermi sea of b-fermions, and they are related to each other by a local spin rotation. However small that may be, any onsite and intraband coupling that breaks the espin local symmetry will select the espin direction for each site and band so that the total energy is lowered.

At time zero, we assume the whole system is in the mixture of these two orthogonal "vacuum" states related by a global espin flip, and then the inter-site kinetic hopping term is turned on adiabatically, the two states $|Gnd\rangle$ and $|\widetilde{Gnd}\rangle$ evolved from these two orthogonal states remain orthogonal, as the Hamiltonian commutes with total espin projection $\Sigma_3$ due to Eq. \eqref{particle espin relation}

In the following, we will present the local basis reduction argument for understanding of the above mathematical constructs \cite{WangCheng2020}.
Any pure state for the total system can be expressed as 
\begin{equation}
    |\Psi\rangle = \sum_{i \mu} C_{i \mu} |i\rangle |\mu\rangle
\end{equation}
where $\{|i\rangle\}$ is the complete orthonormal basis for the subsystem $S$, and $\{|\mu\rangle\}$ is the complete orthonormal basis of subsystem $E$. We apply the Singular Value Decomposition (SVD) to the matrix $C_{i\mu}$, with unitary matrices $U$ and $V$:
\begin{equation}
    C_{i\mu} = \sum_j U_{i j}\rho_j (V^\dagger)_{j\mu}
\end{equation}
such that 
\begin{equation}
    |\Psi\rangle = \sum_j \rho_j |S, j\rangle |E, j\rangle
\end{equation}
where because U is unitary, we have 
\begin{equation}
    |S, j\rangle = \sum_i U_{i j} |i\rangle  \  \ \langle S,i|S,j\rangle = \delta_{i j}
\end{equation}
and because $V$ is unitary, we have
\begin{equation}
    |E, j\rangle = \sum_\mu V_{\mu j} |\mu\rangle  \  \ \langle E,i|E,j\rangle = \delta_{i j}
\end{equation}

Apply the above result to an eigenstate, we have that any eigenstate, ground state in particular, of the full Hamiltonian can always be expressed as 
\begin{equation}
\begin{aligned}
    &|\Psi\rangle =   c_{\Psi, \uparrow}({\bf x})|{\bf x}\uparrow\rangle |E_+\rangle  
    +c_{\Psi, \downarrow}({\bf x}) |{\bf x}\downarrow\rangle |E_- \rangle\\ &+c_{\Psi,0}({\bf x}) |Vac({\bf x})\rangle |E_0\rangle+c_{\Psi, 2}({\bf x})\hat{\xi}^{\dagger}_{\uparrow}({\bf x})\hat{\xi}^{\dagger}_{\downarrow}({\bf x}) |Vac({\bf x})\rangle |E_2 \rangle\\
\end{aligned}
\end{equation}
where all $|E_i\rangle$ are normalized to 1.

The effective single site Hamiltonian for solving the variational coefficients $c_\Psi ({\bf x})$ is
\begin{equation}
    \begin{bmatrix}
    h^{ee} & h^{eo} \\
    h^{oe} & h^{oo}
    \end{bmatrix} =
    \begin{bmatrix}
    h_{0,0} & h_{0,2} & h_{0,\uparrow} & h_{0, \downarrow} \\
    h_{2,0}& h_{2,2} & h_{2, \uparrow} & h_{2, \downarrow}\\
    h_{0,\uparrow}^* & h_{2, \uparrow}^* & h_{\uparrow, \uparrow} & h_{\uparrow, \downarrow}\\
    h_{0, \downarrow}^* & h_{2, \downarrow}^* & h_{\uparrow, \downarrow}^* & h_{\downarrow, \downarrow}
    \end{bmatrix}
\end{equation}

Now the local effective Hamiltonian is ranked no more than 4. we will prove that unless all eigenvalues are degenerate,  effective Hamiltonian for the ground state has to be ranked less than 4, that is the highest energy state will be orthogonal to the ground state and the other two excited states of the local effective Hamiltonian.
The proof is straightforward. Since $-H$ has the exact same eigenstates as $H$, except complete reverse the order of eigenvalues, i.e., ground state for $H$ becomes the highest energy state for $-H$ and vice versa, unless they are the same energy and in that case, all 4 states are degenerate. Thus in a self-consistent solution for an interacting fermion system, we expect that only three local states are needed for variational solutions of ground state and single fermion excitations.

For interacting system, due to the expected symmetry breaking for interacting fermion system, the self-consistent condition for ground state is then
\begin{equation}
    c_2({\bf x}) = 0, \forall {\bf x}\\
\end{equation}
and $c_0({\bf x})$ satisfying
\begin{equation} \label{particle_number_condition}
    \sum_{\bf x} \{1-|c_0({\bf x})|^2 + |c_0({\bf x})|^2n_{Vac}({\bf x})\} = N.
\end{equation}

Thus the self-consistent effective single site Hamiltonian for ground state is reduced to 
\begin{equation} \label{reduced_single_site}
\begin{bmatrix}
    \epsilon_0({\bf x})  & h_{0,\uparrow}({\bf x}) & h_{0, \downarrow}({\bf x}) \\

    h_{0,\uparrow}^*({\bf x})  & h_{\uparrow, \uparrow}({\bf x}) & h_{\uparrow, \downarrow}({\bf x})\\
    h_{0, \downarrow}^* ({\bf x}) & h_{\uparrow, \downarrow}^*({\bf x}) & h_{\downarrow, \downarrow}({\bf x})
    \end{bmatrix}
\end{equation}
The reduction of basis only happens at self-consistency, where the double occupancy of the b-fermions has been renormalized away in the self-consistent vacuum state.

The ground state $|Gnd\rangle$ can be viewed as the resulting adiabatic switching on of the coupling term that gives rise to the off-diagonal terms $h_{0,\sigma}({\bf x})$, i.e., the intersite kinetic hopping terms that exchange fermions between $S$ and its environment. The reduced single site Hamiltonian can be solved analytically. The intersite coupling terms will induce the spontaneous time reversal symmetry breaking, as it mixed the even number particle states with odd number fermionic state at each site ${\bf x}$.

The rest of the paper will discuss the practical implications of the theory.

\section{Constructive Density Functional Theory}
The Ansatz for choosing only three local many-body basis states for a full many body variational reconstruction of the Hamiltonian to an effective Hamiltonian will naturally lead to the quadratic form of the effective Hamiltonian, as we explicit show in this section.
Using the reduced 3 local states $|Vac({\bf x})\rangle, \hat{\xi}_\sigma({\bf x})$, and explicitly consider at each ${\bf x}$, we have multiple bands indexed by $n$, we have the following effective (renormalized) Hamiltonian for the full system, in terms of $\hat{\xi}$ b-fermions:
\begin{equation}
    \hat{H}_{eff} = \sum_{\bf x}\hat{H}_{eff}({\bf x}) + \hat{H}_{eff}^{TB} + \hat{H}_{eff}^{Pairing} +E_{vac}
\end{equation}
where $E_{vac} = \langle Vac| \hat{H}|Vac\rangle$ and

\begin{equation}
\begin{aligned}
    \hat{H}_{eff}({\bf x})=&\sum_{n, \sigma, {\bf x}} \varepsilon_{n\sigma}({\bf x})\hat{\xi}^\dagger_{n\sigma}({\bf x}) \hat{\xi}_{n\sigma} ({\bf x}) + \\
    &\sum_{(n\sigma)\ne (n'\sigma'), {\bf x}} \eta^{\sigma \sigma'}_{n,n'}({\bf x}) \hat{\xi}^\dagger_{n\sigma}({\bf x}) \hat{\xi}_{n'\sigma'}({\bf x})
\end{aligned}
\end{equation}

\begin{equation}
    \hat{H}_{eff}^{TB} =  \sum_{\langle{\bf x}\neq{\bf x'}\rangle; n\sigma n' \sigma'} \lambda^{\sigma \sigma'}_{n n'}({\bf x},{\bf x'}) \hat{\xi}^\dagger_{n\sigma}({\bf x}) \hat{\xi}_{n'\sigma'}({\bf x'}) + h.c.
\end{equation}

\begin{equation} \label{pairing hamiltonian}
\begin{aligned}
    &\hat{H}_{eff}^{Pairing} = \sum_{\langle{\bf x}\neq{\bf x'}\rangle; n\sigma n'\sigma'} \Delta^{\sigma\sigma'}_{n n'}({\bf x},{\bf x'}) \hat{\xi}^\dagger_{n\sigma}({\bf x}) \hat{\xi}^{\dagger}_{n'\sigma'}({\bf x'})    \\ &+\sum_{{\bf x}, n\sigma, n'\sigma'} U_{n,n'}({\bf x}) \xi^\dagger_{n\sigma} ({\bf x})\xi^\dagger_{n'\sigma'} ({\bf x}) \langle \xi_{n'\sigma'} ({\bf x})\xi_{n\sigma} ({\bf x})\rangle   \\
    &+\sum_{{\bf x}, n, n'} J_{n,n'}({\bf x}) \xi^\dagger_{n\uparrow} ({\bf x})\xi^\dagger_{n'\downarrow} ({\bf x}) \langle \xi_{n\downarrow} ({\bf x})\xi_{n'\uparrow} ({\bf x})\rangle \\
    & + h.c.
\end{aligned}
\end{equation}
where the on-site pairing term comes from the on-site 2-particle Coulomb, and exchange interaction terms, and

\begin{equation}
\begin{aligned}
    \varepsilon_{n\sigma}({\bf x}) = &\langle Vac({\bf x})|\hat{\xi}_{n\sigma} ({\bf x}) \hat{H}({\bf x}) \hat{\xi}^\dagger_{n\sigma}({\bf x}) |Vac({\bf x})\rangle \\
    -& \langle Vac({\bf x})| \hat{H}({\bf x}) |Vac({\bf x})\rangle
\end{aligned}
\end{equation}

\begin{equation}
   \eta^{\sigma \sigma'}_{n,n'}({\bf x}) = \langle Vac({\bf x})|\hat{\xi}_{n\sigma} ({\bf x}) \hat{H}({\bf x}) \hat{\xi}^\dagger_{n'\sigma'}({\bf x}) |Vac({\bf x})\rangle 
\end{equation}
and $\lambda^{\sigma \sigma'}_{n,n'}({\bf x}, {\bf x'})$, $\Delta^{\sigma \sigma'}_{n,n'}({\bf x}, {\bf x'})$ are given by the following inter-site hopping term:
\begin{equation}
   t^{\sigma \sigma'}_{n,n'}({\bf x}, {\bf x'}) \hat{\psi}^\dagger_{n\sigma}({\bf x})\hat{\psi}_{n'\sigma'}({\bf x'}) + h.c.
\end{equation}
using the inverse-Bogolyubov transformation
\begin{equation}
\begin{aligned}
    \hat{\psi}_{\downarrow}({\bf x}) &=&\alpha({\bf x}) \hat{\xi}_{\downarrow}({\bf x})+\beta({\bf x})e^{i\varphi({\bf x})} \hat{\xi}^{\dagger}_{\uparrow}({\bf x}) \\
    \hat{\psi}^{\dagger}_{\downarrow}({\bf x}) &=& \alpha({\bf x}) \hat{\xi}^{\dagger}_{\downarrow}({\bf x})+\beta({\bf x})e^{-i\varphi({\bf x})} \hat{\xi}_{\uparrow}({\bf x})\\
    \hat{\psi}_{\uparrow}({\bf x}) &=& \beta({\bf x})e^{i\varphi({\bf x})} \hat{\xi}^{\dagger}_{\downarrow}({\bf x})-\alpha({\bf x}) \hat{\xi}_{\uparrow}({\bf x}) \\
    \hat{\psi}^{\dagger}_{\uparrow}({\bf x}) &=& \beta({\bf x})e^{-i\varphi({\bf x})} \hat{\xi}_{\downarrow}({\bf x})-\alpha({\bf x}) \hat{\xi}^{\dagger}_{\uparrow}({\bf x})
\end{aligned}
\end{equation}

The intersite hopping matrix elements can also be calculated directly as 
\begin{equation}
   \lambda^{\sigma \sigma'}_{n,n'}({\bf x, x'}) = \langle Vac({\bf x})|\hat{\xi}_{n\sigma} ({\bf x}) \hat{H} - \hat{H}({\bf x}) - \hat{H}({\bf x'}) \hat{\xi}^\dagger_{n'\sigma'}({\bf x'}) |Vac({\bf x'})\rangle 
\end{equation}

The effective Hamiltonian of b-fermions is of the generalized quadratic Gor'kov-Bogolyubov-De Gennes \cite{bDG} form 
\begin{equation}
    \hat{H}_{eff} = \sum_{\omega, \mu, \sigma; \omega', \mu', \sigma'} T^{\omega\omega'}_{\mu, \sigma; \mu', \sigma'} \hat{\xi}^{\omega}_{ \mu, \sigma} \hat{\xi}^{ \omega'}_{\mu', \sigma'} + E(\{\beta({\bf x})\})
\end{equation}
where $\omega=\{+/creation, -/annihilation\}$. Note that since $T$ contains the pairing terms, the matrix T is Hermitian, as the Hamiltonian remains to be Hermitian. 
In block presentation of $T$:
\begin{equation} \label{block matrix}
    \begin{bmatrix}
    \Lambda & \Gamma\\
    \Gamma^\dagger & 0
    \end{bmatrix}
\end{equation}
The diagonal blocks remain to be Hermitian, they are the regular tight binding Hamiltonian for b-fermions and the off-diagonal blocks are the pairing terms. 
The Hermitian Hamiltonian can be diagonalized by the following generalized Bogolyubov transformation (mathematically equivalent to Singular Value Decomposition)
\begin{equation} \label{T matrix}
    T=U^\dagger |E|\cdot Sign(E) U
\end{equation}
such that
\begin{equation}
    \hat{H}_{eff}=E(\{\alpha({\bf x}), \beta({\bf x}), \varphi({\bf x})\})+\sum_i \varepsilon_i \hat{\gamma}^\dagger_i\hat{\gamma}_i;   \     \ \varepsilon_i > 0, \  \ \forall i
\end{equation}
where $\varepsilon_i$ are singular values of $T$ and 
\begin{equation}
    \hat{\gamma}_i = \sum_{\omega', \mu', \sigma'}U_{i;\omega', \mu', \sigma'}\hat{\xi}_{\omega', \mu', \sigma'} = 
    u_i\hat{\xi}_{i u} + v_i \hat{\xi}^\dagger_{i d},\  \ i=\{\omega, \mu, \sigma\}
\end{equation}

Here we have effectively derived an exact density functional for total energy~\cite{hohenberg1964pr}  $E(\{\alpha({\bf x}), \beta({\bf x}), \varphi({\bf x})\})$, and it corresponds to the expectation value of the total Hamiltonian with respect to the following pairing ground state, 
\begin{equation}
    |Gnd\rangle =\sum_{i\in\{i| e_i \le 0\}} \hat{\gamma}^\dagger_i |Vac\rangle
\end{equation}
where $e_i$ are the eigenvalues. Note that for Hermitian matrices, single values and eigenvalues are related by taking an absolute value.
The variational total energy functional for the ground state is 
\begin{equation}
    E_{tot} (\{n_k({\bf x}), \varphi_{k \sigma}{\bf x})\}) = \langle Vac | \hat{H} | Vac \rangle + \sum_{\i\in\{i| e_i \le 0\}}  e_i ,  
\end{equation}
where $k$ is the band index.
Note that in this functional, the charge density $n_k({\bf x})$ can be fixed by the self-consistency loop that requires all the filled state gives the expected total charge of the system. For the gauge field $\varphi_k({\bf x}){\bf x}$, which gives rise responsible for bosonic collective excitations (espinons and spinons) in long wavelength limit. It is entirely possible that there are many degenerate $\varphi_i({\bf x})$ configurations for large cluster or large supercell lattices.

The theory is exact in that it not only gives the exact ground state energy functional, it also gives the single fermion excitation spectrum with respect to the ground state specified by the ground state gauge field $\varphi_k({\bf x})$. 
Note that this variational wavefunction is then used to calculate the on-site pairing potential $\langle \xi_{n\uparrow} ({\bf x})\xi_{n\downarrow} ({\bf x})\rangle$ in Equation \eqref{pairing hamiltonian} which is zero when self-consistency is achieved.

We thus can define the following energy gap for single fermion excitation energy as the minimum energy of the Bogolyubov-De Gennes equation, and by the way it is constructed, the gap is a functional of the U(1) gauge field $\varphi({\bf x})$, and the charge density field $2\beta^2({\bf x})$. 

The self-consistency condition for $2\beta^2({\bf x})$ ensures that the the ground state generated b-fermions condensate charge density is consistent with the doping level, i.e., the charge density. 
Note that the algorithm is polynomial in number of sites $N_x$, as opposed to the full diagonalization of a full many-body Hamiltonian which scales exponentially with $N_x$. Furthermore, we can use the traditional LDA for an initial guess of $\beta({\bf x})$, and calculate the tight binding hopping parameters from first principles LDA calculation, which will further improve the scaling performance to $O(N)$.

Before we move on to some  example applications of the theory, we want to point out that the $\varphi({\bf x})$ phase can be absorbed in $\hat{\psi}^\dagger$ operators. This would bring the dependency on $\varphi({\bf x})$ to the hopping matrix elements of the Hamiltonian. For dimer solution below, this turned out to be a convenience.

\section{Low frequency resonance in water due to off-diagonal topological order of protons}
For water, for each water molecule, we map the proton states into the following local states
\begin{equation}
\begin{aligned}
    &|Vac({\bf x})\rangle = |H_2O\rangle, \\ &|p_\sigma({\bf x})\rangle = \hat{p}_\sigma^\dagger({\bf x})|Vac({\bf x})\rangle = |H_3O^+_\sigma\rangle, \\  &|h_\sigma({\bf x})\rangle =  \hat{h}^\dagger_\sigma({\bf x})|Vac({\bf x})\rangle =  \hat{p}_{-\sigma}({\bf x})|H_2O\rangle = |HO^-_\sigma\rangle
\end{aligned}
\end{equation}
where ${\bf x}$ is the center of charge of the $H_2O$ molecular.
Then we have the following effective quadratic Hamiltonian for protons, with the kinetic hopping terms has the built-in disorder due to the nature of the water being liquid.
\begin{equation} \label{water effective hamiltonian}
\begin{aligned}
    \hat{H}_{eff} =& \sum_{{\bf x} \sigma} \{\varepsilon_p \hat{p}_\sigma^\dagger({\bf x}) \hat{p}_\sigma({\bf x})   + \varepsilon_h \hat{h}_\sigma^\dagger({\bf x}) \hat{h}_\sigma({\bf x})\}  \\ &+\sum_{\bf x x'} t({\bf x, x'}) \hat{p}_\sigma^\dagger({\bf x}) \hat{h}_{-\sigma}^\dagger({\bf x'}) + h.c.
\end{aligned}
\end{equation}
The diagonal single particle energy for protons and holes are the same for pure water and they may differ when the number of protons and number of holes differ, resulting in the pH level deviates from 7. When spin-orbit coupling is considered, the two degenerate state due to parity conservation will be split and the energy splitting between the two levels is then $\sqrt{\zeta_{SO}^2 + w(pH-7)^2}$ with $w$ a positive scaling factor.

The eigenvalues $E$ of Hamiltonian Eq. \eqref{water effective hamiltonian} satisfy
\begin{equation}
    \sum_{\bf x'} \begin{bmatrix}
    (\varepsilon_p-E) \delta({\bf x, x'}) & t({\bf x, x'})\\
    t^\dagger({\bf x, x'}) & (\varepsilon_h -E)\delta({\bf x, x'})
    \end{bmatrix}\cdot\begin{bmatrix}
   {\bf u_p(x')}\\ {\bf u_h(x') }
    \end{bmatrix} = 0
\end{equation}
which can be reduced to
\begin{equation}
    \{T^\dagger T - (\varepsilon_p -E)(\varepsilon_h -E)\} \cdot {\bf u_h} = 0
\end{equation}
Thus the eigenvalues of Eq. \eqref{water effective hamiltonian} are the roots of the following polynomial
\begin{equation}
    \prod_i (t_i^2 - (\varepsilon_p -E)(\varepsilon_h -E)) = 0
\end{equation}
where $t_i^2$ are the eigenvalues of the matrix $T^\dagger T$. 

The single fermion eigenvalues are then solved as 
\begin{equation}
    E_{i\pm} = \frac{\varepsilon_p+\varepsilon_h\pm\sqrt{(\varepsilon_p-\varepsilon_h)^2 + 4t_i^2 } }{2}
\end{equation}
with he chemical potential in the middle 
\begin{equation}
    \mu = \frac{\varepsilon_p+\varepsilon_h}{2}
\end{equation}

This energy spectrum is completely determined by the energy spectrum of $T^\dagger T$, and the gap of the proton-hole pair $\Delta=|\varepsilon_p-\varepsilon_h|$, which determines the density of the hydrogen bond. 

In our theory, the topological order of water is originated from the high density of states at zero eigenvalues of $T^\dagger T$. Note that the range of the eigenvalues is bounded by the largest nearest neighbor hopping, due to the long range disorder, the density of states for $T^\dagger T$ will diverge at $t^2=0$ in the thermodynamic limit, leading to topological order. Without long range disorder, the density of state at zero energy will not diverge and the topological order will be quenched. This can be more rigorously shown by noting that the self-consistent solution requires the gradient with respect to the complex variable $\beta_n({\bf x})e^{i\varphi_n({\bf x})}$ to be zero, thus implying the compatibility of periodicity for the charge order $\beta_n({\bf x})$ and topological order $\varphi_n({\bf x})$. For $H_2O$, for example, below freezing temperature, crystalline ice forms and the topological order shall be expected to be destroyed. 

Under the topological order phase (liquid water), and non-zero $\varepsilon_p-\varepsilon_h$, we shall conclude to have two sharp peaks in single fermion excitation density of states, one below the chemical potential and one above. As a consequence, a sharp electrical resonant absorption is expected at the frequency $\Delta/\hbar$. We do expect $\Delta$ to change with relative pH level of the water, for pH$=7$, we expect to have the minimum $\Delta$, which is the size of spin-orbit coupling of the hopping proton at the local hydrogen bond. In other words, the excitations for pH$=7$, the Majorana fermions gain mass due to spin-orbit coupling, another topological effect found similarly for topological insulators~\cite{hasan2010rmp} in electronic systems. 
$\Delta$ increases with $|pH-7|$. The mass of the Majorana fermion is not temperature sensitive and the excitation is considered elementary particle-like, i.e., the corresponding resonant width is extremely sharp and the resonance has a extremely long lifetime. Using the estimate of spin-orbit coupling of electrons of $~1 meV$ and scale it with the proton mass factor $~1/1840$, we get the Majorana fermion mass of $0.5 \mu eV$, which put it at $~100 MHz$ range.

Thus $\Delta$ is expected to be very sensitive to local pH levels in a heterogeneous aqueous system, as well as the the details in the hydrogen bond and the strength of spin-orbit coupling of protons at the hydrogen bond.   

Our theoretical framework opened the door to treat the large bio-molecules in aqueous solutions. An important observation is that although the off-diagonal topological ordering are very important in maintaining the long life times of resonances, the channel specificity of these resonances is due to the local energy splitting, e.g., for hydrogen bonds.

\section{Hubbard Model Example}
For asymmetric Hubbard dimer, and the lattice version of it, without spin-orbit coupling, applying the theory, we have the single b-fermion excitation spectrum can be solved with the following quadratic form of Hamiltonian
\begin{equation}
    \hat{H}_{eff} = {\bf \hat{v}}^\dagger \begin{bmatrix}
    H_{AB} & 0\\
    0 & H_{BA}
    \end{bmatrix} {\bf \hat{v}}
\end{equation}
where ${\bf \hat{v}}^\dagger$ is the complex conjugate of the column vector operator ${\bf \hat{v}}$, given, e.g., for dimer and periodic lattice of the dimer as
\begin{equation}
    {\bf \hat{v}}^\dagger = \begin{bmatrix}
    \hat{\xi}^\dagger_{A\uparrow} &\hat{\xi}_{A\downarrow} & \hat{\xi}^\dagger_{B\uparrow} &\hat{\xi}_{B\downarrow} & \hat{\xi}^\dagger_{B\downarrow} &\hat{\xi}_{B\uparrow} & \hat{\xi}^\dagger_{A\downarrow} &\hat{\xi}_{A\uparrow}
    \end{bmatrix}
\end{equation}
Note that if spin-orbit coupling is zero, the two diagonal blocks are degenerate due to parity symmetry. Otherwise, we would have a spin-orbit coupling split two bands for the single b-fermion excitation that do not couple to each other. In the thermodynamic limit, these two bands correspond to two different boundary condition or edge states with different chirality. Without loss of generality, we will discuss the excitation spectrum of one of them here.

Assume we have reached the self-consistency for charge condition so that $\beta({\bf x})$ are fixed by the chemical potential and the asymmetry between the two dimer site are captured by the onsite energy differences of the self-consistent b-fermions, we would have for the reduce effective Hamiltonian $H_{AB}$, the following general form
\begin{equation}
H_{AB} = \begin{bmatrix}
|\varepsilon_A| sign(A) & T \\
T^\dagger & |\varepsilon_B| sign(B)
\end{bmatrix},
\end{equation}
where $\varepsilon_A, \varepsilon_B$ are diagonal but alternate in sign.
The singular value of the above matrix are the square root of the eigenvalues of the following non-negative definite matrix
\begin{equation}
    H_{AB}H_{AB}=\begin{bmatrix}
    \varepsilon_A^2 + T T^\dagger & \Delta \\
    \Delta^\dagger & \varepsilon_B^2 + T^\dagger T
    \end{bmatrix},
\end{equation}
where 
\begin{equation}
    \Delta = \varepsilon_A T^\dagger + T \varepsilon_B
\end{equation}
Define the following Hermitian matrix
\begin{equation}
    \mathcal{T} = T T^\dagger = U^\dagger_{\mathcal{T}} E^2_{\mathcal{T}} U_{\mathcal{T}}
\end{equation}
we have 
\begin{equation}
    sign(E^2_{\mathcal{T}}) = I
\end{equation}
i.e., all eigenvalues of $\mathcal{T}$ are non-negative, and they are all positive when spin-orbit coupling is not zero
\begin{equation}
    E_{\mathcal{T}} = \sqrt{E^2_{\mathcal{T}}(\zeta_{SO}=0) + w \zeta^2_{SO}}
\end{equation}
We then have 
\begin{equation}
    T^\dagger T = U_{\mathcal{T}}  E^2_{\mathcal{T}} U^\dagger_{\mathcal{T}}
\end{equation}
So we can perform the following unitary transformation to $H^2_{AB}$ as
\begin{equation}
    \tilde{H}^2_{AB} = \begin{bmatrix}
    U^\dagger_{\mathcal{T}}  & 0 \\
    0 &  U_{\mathcal{T}}
    \end{bmatrix} H^2_{AB} \begin{bmatrix}
     U_{\mathcal{T}} & 0 \\
    0 & U^\dagger_{\mathcal{T}}
    \end{bmatrix} 
\end{equation}
which has the same eigenvalues and singular values as $H^2_{AB}$.
Explicitly $\tilde{H}^2_{AB}$ is 
\begin{equation}
    \tilde{H}^2_{AB}=\begin{bmatrix}
    \varepsilon^2_A  +  E^2_{\mathcal{T}} & \tilde{\Delta}\\
    \tilde{\Delta}^\dagger  & \varepsilon^2_B  +  E^2_{\mathcal{T}} 
    \end{bmatrix} 
\end{equation}
The singular values of the matrix $\tilde{H}^2_{AB} - (E^2_{\mathcal{T}} + \frac{\varepsilon_A^2 + \varepsilon_B^2}{2})\bigotimes I_2 $, $\mathcal{E}^2_\kappa$, is then solved using the following
\begin{equation}
    \mathcal{E}^2_\kappa = \sqrt{(\tilde{\Delta}^\dagger\tilde{\Delta})_\kappa + (\varepsilon_A^2-\varepsilon_B^2)^2}
\end{equation}
where $(\tilde{\Delta}^\dagger\tilde{\Delta})_\kappa$ are the $\kappa^{th}$ singular value of the matrix $\tilde{\Delta}^\dagger\tilde{\Delta}$
So the excitation spectrum of the 
$H_{AB}$ is then 
\begin{equation}
    \sqrt{E^2_{\mathcal{T}} + \frac{\varepsilon_A^2 + \varepsilon_B^2}{2} + \mathcal{E}^2_\kappa}
\end{equation}
That is, the single fermion excitation always has a gap, as long as $\mathcal{E}^2_\kappa >0$.

At symmetric case, the Hubbard model has been proposed as the simplest model for high-Tc superconductors and the Anti-ferromagnetic Neel phase at zero doping. An early phenomenological model \cite{so5theory} points out the role of doping can be viewed as an mechanism for breaking the $SO(5)$ symmetry, driving the system from AFM to SC. Our theory is consistent with these conjectures. On the other hand, we think to ultimately settle down on the theory of HTC, an application of our theory to the real materials is needed.

\section{Conclusion}
It is clear that the problems and systems the proposed theory can be applied to are not limited to the examples presented in this paper. Most importantly from the computational perspective, the existing enterprise on first principles Local Density Functional electronic calculations provides an enormous leverage and starting point for our theory as it provides an excellent starting charge density for each band. On the theoretical front, since the theory is based on the eigenstate solutions of a Hamiltonian, the bosonic collective excitations of the Hamiltonian will be studied as the excitations of the phase field $\varphi_n({\bf x})$, or gauge fields. The rich Lee-group symmetry structure offered by going beyond single band model, namely p-band, d-band, f-band and beyond, and the additional espin $U(2)\bigoplus AU(2)$ symmetry coupling to spin $U(2) \bigoplus AU(2)$ symmetry will also be explored.

{\bf Author contributions} Xindong Wang is responsible for sections 1 and 2, all authors contributed to the applications of the theory, sections 3 and 4.

{\bf Acknowledgement}
X.D. Wang acknowledge Sophyics Technology, LLC. H.-P Cheng and X. Chen acknowledge US Department of Energy (DOE), Office of Basic EnergySciences (BES), Contract No. DE-FG02-02ER45995. 

{\bf Corresponding author} xdwang@sophyicstech.com

\bibliography{./refs.bib}

\begin{thebibliography}{1}
\expandafter\ifx\csname url\endcsname\relax
  \def\url#1{\texttt{#1}}\fi
\expandafter\ifx\csname urlprefix\endcsname\relax\def\urlprefix{URL }\fi
\providecommand{\bibinfo}[2]{#2}
\providecommand{\eprint}[2][]{\url{#2}}

\bibitem{hohenberg1964pr}
\bibinfo{author}{Hohenberg, P.} \& \bibinfo{author}{Kohn, W.}
\newblock \bibinfo{title}{Inhomogeneous electron gas}.
\newblock \emph{\bibinfo{journal}{Phys. Rev.}} \textbf{\bibinfo{volume}{136}},
  \bibinfo{pages}{B864--B871} (\bibinfo{year}{1964}).
\newblock \urlprefix\url{https://link.aps.org/doi/10.1103/PhysRev.136.B864}.

\bibitem{Lowdin1966pr}
\bibinfo{author}{L{\"o}wdin, P.-O.}
\newblock \bibinfo{title}{Aperiodic solid: some aspects on the biological
  problems of heredity, mutations, aging, and tumors in view of the quantum
  theory of the dna molecule}.
\newblock \emph{\bibinfo{journal}{Advance in Quantum Chemistry}}
  \textbf{\bibinfo{volume}{2}}, \bibinfo{pages}{213--360}
  (\bibinfo{year}{1966}).

\bibitem{WangCheng2020}
\bibinfo{author}{Wang, X.} \& \bibinfo{author}{Cheng, H.-P.}
\newblock \bibinfo{title}{Self-consistent effective hamiltonian theory for
  fermionic many body systems}.
\newblock \emph{\bibinfo{journal}{International Journal of Modern Physics}}
  (\bibinfo{year}{2020}).
\newblock \bibinfo{note}{Accepted for publication}.

\bibitem{bDG}
\bibinfo{author}{De~Gennes, P.} \& \bibinfo{author}{Pincus, P.}
\newblock \emph{\bibinfo{title}{Superconductivity Of Metals And Alloys}}.
\newblock Advanced Books Classics (\bibinfo{publisher}{Avalon Publishing},
  \bibinfo{year}{1999}).
\newblock \urlprefix\url{https://books.google.com/books?id=xacsAAAAYAAJ}.

\bibitem{hasan2010rmp}
\bibinfo{author}{Hasan, M.~Z.} \& \bibinfo{author}{Kane, C.~L.}
\newblock \bibinfo{title}{Colloquium: Topological insulators}.
\newblock \emph{\bibinfo{journal}{Rev. Mod. Phys.}}
  \textbf{\bibinfo{volume}{82}}, \bibinfo{pages}{3045--3067}
  (\bibinfo{year}{2010}).
\newblock \urlprefix\url{https://link.aps.org/doi/10.1103/RevModPhys.82.3045}.

\bibitem{so5theory}
\bibinfo{author}{Demler, E.}, \bibinfo{author}{Hanke, W.} \&
  \bibinfo{author}{Zhang, S.-C.}
\newblock \bibinfo{title}{So(5) theory of antiferromagnetism and
  superconductivity}.
\newblock \emph{\bibinfo{journal}{Rev. Mod. Phys.}}
  \textbf{\bibinfo{volume}{76}}, \bibinfo{pages}{909--} (\bibinfo{year}{2004}).
\newblock \urlprefix\url{https://doi.org/10.1103/RevModPhys.76.909}.

\end{thebibliography}

\end{document}